\documentclass[12pt]{article}
\usepackage{a4,graphicx}
\usepackage[small,bf]{caption}

\def\lsim{\mathrel{\rlap{\lower4pt\hbox{\hskip1pt$\sim$}}
    \raise1pt\hbox{$<$}}}                
\def\gsim{\mathrel{\rlap{\lower4pt\hbox{\hskip1pt$\sim$}}
    \raise1pt\hbox{$>$}}}                

\newcommand{\EM}{\mbox{\tiny $\rm E\!M$}}                

\begin{document}

\title{
\vspace{-3.25cm}
\flushright{\small ADP-15-22/T924} \\
\vspace{-0.35cm}
{\small Edinburgh 2015/10} \\
\vspace{-0.35cm}
{\small Liverpool LTH 1048} \\
\vspace{-0.35cm}
{\small July 28, 2015} \\
\vspace{0.5cm}
\centering{\Large \bf Reply to ``Comment on `Lattice determination of
           Sigma -- Lambda mixing' ''}}

\author{\large
         R. Horsley$^a$, J. Najjar$^b$, Y. Nakamura$^c$, H. Perlt$^d$, \\
         D. Pleiter$^e$, P.~E.~L. Rakow$^f$, G. Schierholz$^g$, \\
         A. Schiller$^d$, H. St\"uben$^h$ and J.~M. Zanotti$^i$ \\[1em]
         -- QCDSF-UKQCD Collaboration -- \\[1em]
        \small $^a$ School of Physics and Astronomy,
               University of Edinburgh, \\[-0.5em]
        \small Edinburgh EH9 3FD, UK \\[0.25em]
        \small $^b$ Institut f\"ur Theoretische Physik,
               Universit\"at Regensburg, \\[-0.5em]
        \small 93040 Regensburg, Germany \\[0.25em]
        \small $^c$ RIKEN Advanced Institute for
               Computational Science, \\[-0.5em]
        \small Kobe, Hyogo 650-0047, Japan \\[0.25em]
        \small $^d$ Institut f\"ur Theoretische Physik,
               Universit\"at Leipzig, \\[-0.5em]
        \small 04109 Leipzig, Germany \\[0.25em]
         \small$^e$ J\"ulich Supercomputer Centre,
               Forschungszentrum J\"ulich, \\[-0.5em]
        \small 52425 J\"ulich, Germany \\[0.25em]
        \small $^f$ Theoretical Physics Division,
               Department of Mathematical Sciences, \\[-0.5em]
        \small University of Liverpool,
               Liverpool L69 3BX, UK \\[0.25em]
        \small $^g$ Deutsches Elektronen-Synchrotron DESY, \\[-0.5em]
        \small 22603 Hamburg, Germany \\[0.25em]
        \small $^h$ Regionales Rechenzentrum, Universit\"at Hamburg, \\[-0.5em]
        \small 20146 Hamburg, Germany \\[0.25em]
        \small $^i$ CSSM, Department of Physics,
               University of Adelaide, \\[-0.5em]
        \small Adelaide SA 5005, Australia}

\date{}

\maketitle



\begin{abstract}
In this Reply, we respond to the above Comment. Our
computation [Phys.\ Rev.\ D 91 (2015) 074512] only took
into account pure QCD effects, arising from quark mass differences,
so it is not surprising that there are discrepancies in isospin splittings
and in the Sigma -- Lambda mixing angle. We expect that these discrepancies
will be smaller in a full calculation incorporating QED effects.
\end{abstract}


\clearpage






    We agree with Gal \cite{Gal:2015iha} that lattice numbers
 for the mixing angle are not complete until numbers for the isospin
 splittings are correctly accounted for. In our paper, \cite{SigLam}
 the calculation only includes QCD effects, arising from quark
 mass differences. For a complete result, a full calculation with
 QED effects added is also needed. 


    Before addressing the main issue, we would first like to mention
 that it is not surprising that our results satisfy the Coleman--Glashow
 \cite{coleman61a} and Dalitz--von-Hippel (eq.~(3) of \cite{DvH}) relations.
 Our fit function has all the $SU(3)$ symmetry constraints built in,
 so it automatically obeys every symmetry relation to the appropriate order. 
 In this case, violations of the Coleman--Glashow relation are at
 $O(\delta m^3)$, while violations of the Dalitz--von-Hippel relation
 are at $O(\delta m^2)$. 

\begin{figure}[htbp]
   \begin{center}
      \includegraphics[width=9.70cm]{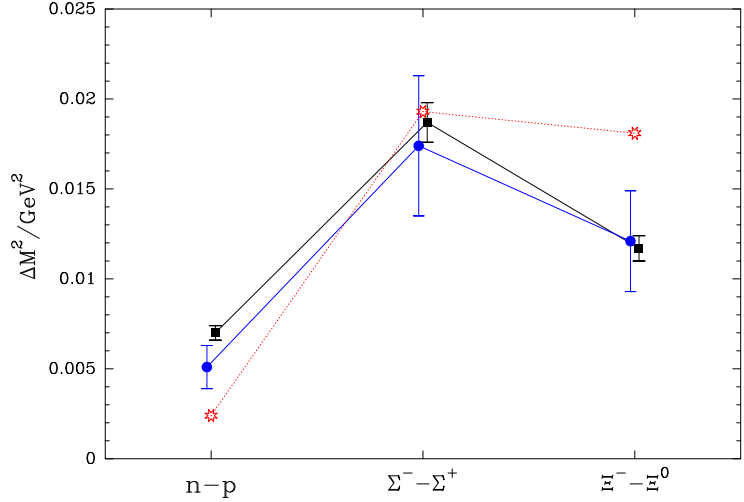}
   \end{center}
\caption{Baryon octet $(\mbox{mass splittings})^2$. 
         Black squares are from eq.~(\ref{splits}), while
         blue circles the lattice QCD results. Both use the 
         pure QCD numbers from table 5, \cite{SigLam}.
         Red stars use the physical values (i.e.\ for QCD and QED).}
\label{splitting_fig}
\end{figure}

    We believe that lattice QCD gives reliable numbers 
 for the part of the isospin splitting due to quark mass differences, 
 and that the differences between the values in~\cite{SigLam}
 and experiment are mainly due to electromagnetic effects. 
    Our reasons to believe that lattice numbers for the purely QCD
 contribution are accurate are: 
   firstly the lattice gives good values for the splittings between 
 the multiplets ($N, \Lambda, \Sigma, \Xi$), and it is 
 hard to see how there could be a systematic error that would
 spoil the isospin splittings without also showing up in 
 splittings between multiplets. 
   Secondly at leading order, pure QCD relates the isospin splittings 
 to the splittings between the multiplets
 through the relations
 \begin{eqnarray} 
 M_n^2 - M_p^2 &\approx&  
 \frac{  m_d -  m_u}{ m_s -  m_u}
  ( M_{\Xi^0}^2 - M_{\Sigma^+}^2) \nonumber \\
 M_{\Sigma^-}^2 - M_{\Sigma^+}^2 &\approx&  
 \frac{ m_d -  m_u}{ m_s - m_u}
  ( M_{\Xi^-}^2 - M_p^2)  \label{splits}\\
 M_{\Xi^-}^2 - M_{\Xi^0}^2 &\approx&  
 \frac{  m_d -  m_u}{ m_s -  m_u}
  (  M_{\Sigma^-}^2 - M_n^2) \nonumber
 \end{eqnarray}
 (see Fig.~2 of \cite{isospin}). Our simulations show that this leading
 order formula only has minor corrections from higher order effects, 
 and the above relations hold reasonably well, with a value $\approx 0.022$ 
 for the quark mass ratio $(m_d-m_u)/(m_s - m_u)$. The main 
 systematic uncertainty in this mass ratio are due to 
 the difficulty of correcting for electromagnetic effects
 in the pseudoscalar meson sector. In Fig.~\ref{splitting_fig}
we show the splitting values. The squares and circles are consistent
with each other, but do not reproduce the full result without adding a 
non-QCD force (namely QED).

     The relations~(\ref{splits}) do not hold in the real world, 
 we take this as evidence that the QED corrections to the 
 isospin splittings are substantial. 
    Simulations with QED included~\cite{Borsanyi}
  show that QED effects on the 
 isospin splittings are comparable with the effects of the
 $(m_d-m_u)$ difference, and that the combined simulation 
 reproduces the physical numbers very closely. Unfortunately 
 we do not have lattice results for the QED shift in the 
 $\Sigma^0$ mass, so we can not estimate the QED effect in 
 the mixing matrix element. 


   In the paper by Isgur \cite{Isgur} there are electrostatic (Coulomb)
 contributions to the isospin splittings, but they cancel completely
 for the combination $[(M_{\Sigma^0} - M_{\Sigma^+}) - (M_n - M_p)]$, 
 which, by the Dalitz--von-Hippel relation, is proportional to 
 the $\Sigma$-$\Lambda$ mixing angle.  If this holds, we would expect
 to see some shifts in the isospin splittings, but  no Coulomb
 contribution to the mixing angle. 
     However, this exact cancellation is model-dependent. The Coulomb
 contributions in \cite{Isgur} are calculated for a particular
 Gaussian wavefunction, which has more symmetry than required by
 QCD. For equal-mass quarks the Isgur spatial wavefunction is completely
 symmetric under exchange of any quark pair. Since the octet is 
 a mixed symmetry multiplet, we do not expect complete symmetry. 
 In the proton ($uud$) there is no theorem that says
 $\langle 1/r_{uu} \rangle = \langle 1/r_{ud} \rangle.$


    We have applied our flavour analysis to QED effects too.
 We find that there are five allowed coefficients for the 
 electromagnetic effects, one of which just shifts all masses
 by the same amount, so making no contribution to splitting
 or mixing. In terms of these coefficients, we find 
 (assuming small mixing angle) 
  \begin{eqnarray}
 M^2_n &=& \frac{1}{3} \left[ 2 B_0^{\EM} + 2 B_1^{\EM}
 + B_2^{\EM} + 3 B_3^{\EM} \right] e^2 \nonumber\\
 M^2_p &=& \frac{1}{3} \left[ 2 B_0^{\EM} + 3 B_1^{\EM}
 - B_2^{\EM} + 3 B_3^{\EM} \right] e^2 \nonumber\\
 M^2_{\Sigma^-} &=& \frac{1}{3} \left[ 2 B_0^{\EM} + B_1^{\EM}
 \right] e^2  \nonumber \\
 M^2_{\Sigma^0} &=& \frac{1}{12} \left[ 8 B_0^{\EM} + 8 B_1^{\EM}
 - 2 B_2^{\EM} + 3 B_3^{\EM} + 9 B_4^{\EM} \right] e^2 \\
 M^2_{\Lambda^0} &=& \frac{1}{12} \left[ 8 B_0^{\EM} + 8 B_1^{\EM}
 + 2 B_2^{\EM} + 9 B_3^{\EM} + 3 B_4^{\EM} \right] e^2  \nonumber \\
 M^2_{\Sigma^+} &=& \frac{1}{3} \left[2 B_0^{\EM} + 3 B_1^{\EM}
 - B_2^{\EM}  + 3 B_3^{\EM} \right] e^2 \nonumber\\
 M^2_{\Xi^-} &=& \frac{1}{3} \left[ 2 B_0^{\EM} + B_1^{\EM}
 \right] e^2\nonumber\\
 M^2_{\Xi^0} &=& \frac{1}{3} \left[ 2 B_0^{\EM} + 2 B_1^{\EM}
 +  B_2^{\EM} + 3 B_3^{\EM} \right] e^2\nonumber
 \end{eqnarray} 
 as the general expression for the electromagnetic 
 contribution to the masses, and 
 \begin{equation} 
 \langle \Sigma^0 | M^2_{\EM} | \Lambda \rangle = 
 \frac {\sqrt{3}}{12} 
 \left[ -2 B^{\EM}_2 - 3 B^{\EM}_3 + 3 B^{\EM}_4 \right ] e^2 
 \label{EMmix} 
 \end{equation} 
 for the mixing matrix element.
 (We fit to the squares of the masses -- the form of the expansion
 is of course the same if a fit to the masses themselves is made.) 
 It may readily be checked that these electromagnetic contributions
 satisfy the Coleman--Glashow and Dalitz--von-Hippel relations,
 for any value of the $B^{\EM}$ coefficients. 

    In the Isgur model, these coefficients are not all 
 independent. The Coulomb terms of the Isgur model 
 follow the pattern 
 \begin{equation} 
 B_3^{\EM} = B_4^{\EM} = - B_1^{\EM} \; , \qquad 
 B_2^{\EM} = 0 
 \end{equation} 
 which in turn ensures that there is no Coulomb 
 contribution to the mixing, see eq.~(\ref{EMmix}). However, these 
 inter-relations are model dependent, and we need 
 lattice calculations to see how well they hold. 


      We are currently working on a combined simulation with 
 QED effects included. Preliminary results suggest that QED 
 effects do account for much of the difference between the 
 current QCD-only results and the experimental values.
    In the Isgur model it is expected that although the 
 individual splittings will be changed by QED effects, 
 the QED contribution to the $\Sigma$-$\Lambda$ mixing angle
 will cancel. This is however a model dependent statement, 
 with other choices for the spatial wavefunction the 
 cancellation would not be complete. 
    Our joint QED and QCD results are not yet at the point
 where we can comment on how much the mixing angle
 is changed by QED, but we are grateful to Gal for bringing
 the issue to our attention, and will certainly include 
 a discussion of the question when we have our final results.


\section*{Acknowledgements}


The numerical configuration generation (using the BQCD lattice
QCD program \cite{nakamura10a}) and data analysis 
(using the Chroma software library \cite{edwards04a}) was carried out
on the IBM BlueGene/Q using DIRAC 2 resources (EPCC, Edinburgh, UK),
the BlueGene/P and Q at NIC (J\"ulich, Germany), the
SGI ICE 8200 and Cray XC30 at HLRN (The North-German Supercomputer
Alliance) and on the NCI National Facility in Canberra, Australia
(supported by the Australian Commonwealth Government).
This investigation has been supported partly 
by the EU Grants No. 227431 (Hadron Physics2) and No. 283826
(Hadron Physics3). JN was partially supported by EU grant 228398
(HPC-EUROPA2). HP was supported by DFG Grant No. SCHI 422/9-1.
PELR was supported in part by the STFC under contract ST/G00062X/1
and JMZ was supported by the Australian Research Council Grant
No. FT100100005 and DP140103067. We thank all funding agencies.



\end{document}